# The future of VLBI

**Huib Jan van Langevelde[1]**
*The Joint Institute for VLBI in Europe*
*Postbus 2, 7990 AA Dwingeloo, the Netherlands*
*And Sterrewacht Leiden, Leiden University*
*Postbus 9513, 2300 RA Leiden, the Netherlands*
E-mail: `langevelde@jive.nl`

Almost two decades after the establishment of the Joint Institute for VLBI in Europe (JIVE), the European VLBI Network is a thriving scientific infrastructure with a significant user community and a healthy proposal pressure. It offers opportunities to address a breadth of important scientific topics, which feature in national and European astronomy roadmaps. Most of these science themes call for further enhancements of the sensitivity and image quality delivered by VLBI networks. The exceptional progress of e-VLBI over the last five years demonstrates how sensitive VLBI should be done in the future. At the same time JIVE is pushing the technology for large capacity correlators that can connect VLBI networks with many elements in real-time. Indeed, many new initiatives to build or outfit telescopes for VLBI are emerging from around the world. The technological VLBI developments have a great synergy with the SKA preparations. This is recognized in the SKA pathfinder role that e-VLBI has in the European VLBI Network for exploring connectivity and real-time science techniques. Moreover, VLBI with its locally visible elements offers great possibilities for training and outreach. With the new scientific capabilities, especially those complementary to other SKA pathfinders, VLBI will be a flourishing scientific instrument for the future.

[1] Speaker

## 1. Background

When JIVE was established as a correlator centre, one of its main characteristics was the ability to play back tape reels from 16 stations simultaneously [1]. A major improvement to the reliability, flexibility and efficiency occurred ten years ago, when hard-disk recording replaced the tape reels. Then VLBI underwent another revolution by providing real-time results, thereby enabling new science, boosting the reliability further and potentially enhancing the sensitivity. This provides a natural growth path for this branch of high-resolution radio astronomy. The increasing quality of the science return of VLBI has led to a solid user interest in the technique. Moreover, it has been demonstrated [6] that VLBI can enhance the scientific impact of space missions, in planetary explorations or fundamental physics. Probably related to this, there are many national initiatives to establish new local VLBI elements.

Radio astronomers around the world are developing the Square Kilometre Array (SKA) [12], which will take radio astronomy into a completely different sensitivity domain. The concerted, global effort has focused on the technical development necessary for the SKA, which has fed into two precursor installations, currently under construction at the proposed SKA sites. Most of the precursor science will focus on surveys with arcsecond scale resolution. Additionally, during the first phase of the SKA the scientific focus is thought to be on transients, pulsars and HI from distant galaxies, requiring low frequency receptors at modest baselines to optimize brightness sensitivity. The scientific strength of VLBI will be to provide capabilities for high-resolution follow-up for all these instruments, requiring sensitive and flexible networks that operate in a global collaboration.

Below I will discuss these aspects in detail and focus on the next steps on this roadmap to the VLBI of the future, a development that will be synergetic with the technical development of the SKA. In recognition of the special nature of this conference, I have often taken a European VLBI Network (EVN[2]) and Joint Institute for VLBI in Europe (JIVE) biased perspective in this discussion.

## 2. Scientific motivation

The EVN is developing a long-term roadmap for the next 5 to 15 years. There are a number of initiatives to define scientific priorities on national and European scales and the VLBI community must set its priorities in view of the plans of radio astronomers all over the world. These efforts continue to be based on the EVN science vision that was developed a couple of years ago: EVN2015 (http://www.evlbi.org/publications/publications.htm).

The EVN science case is addressing a wide range of astronomical topics, starting from the classical area of jets in Active Galactic Nuclei (AGN) and the AGN/starburst connection in cosmological fields, which calls for a boost in sensitivity. More capabilities at higher frequency are also in demand, for example to probe jet physics close to the black hole event horizon. An extreme example of this is the deployment of sub-mm VLBI with ALMA, which promises to map the shadow of the central Galactic black hole, SgrA*. The research in masers will be stimulated by more coverage at 22GHz and 43GHz, offering a view on the dynamics on the small scales that are relevant for star and planet formation. From the same observations usually also very fundamental distance measurements are obtained (Figure 1).

---

[2] The European VLBI Network is a joint facility of European, Chinese, South African and other radio astronomy institutes funded by their national research councils

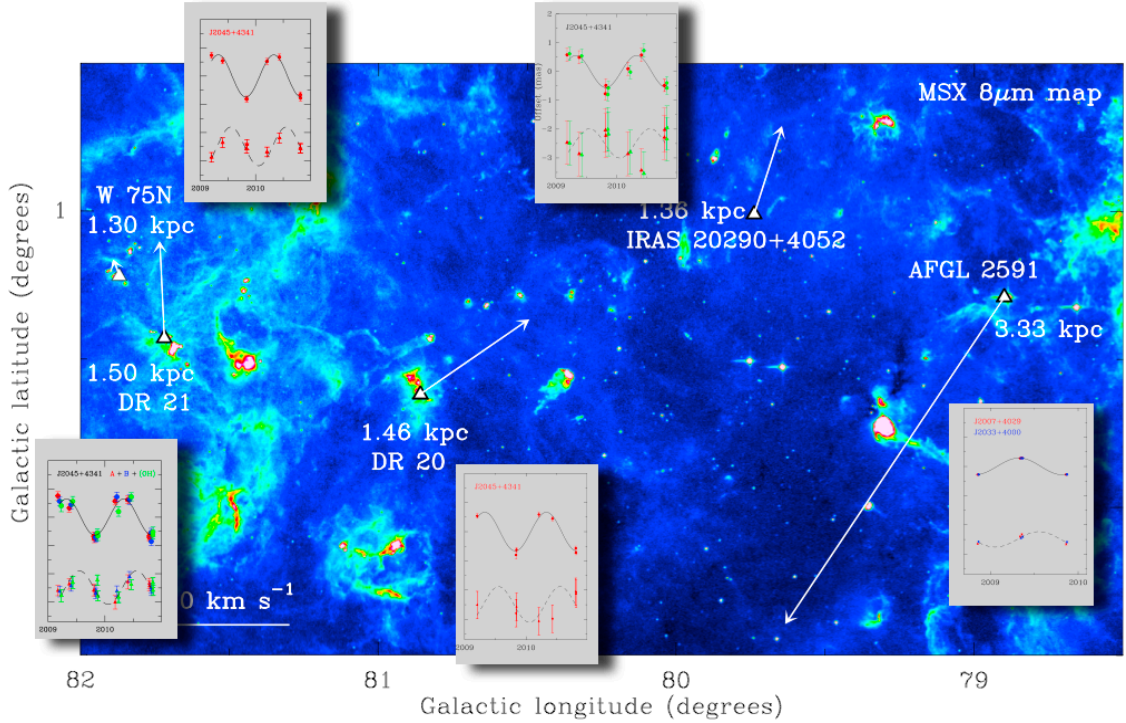

*Figure 1. Parallax measurements of star-forming regions in the Cygnus X region. These EVN observations demonstrate that not all the masers are at the same distance, and do not form a single spatial structure [9]*

In Galactic astronomy, the study of transient radio sources has already taken advantage of the development of real-time VLBI [8], but currently the samples under study are limited in size, until much better sensitivity becomes available. Together with large field-of-view survey instruments, like LOFAR [11], providing many more triggers, this can be expected to be a booming industry in a matter of years. Pulsar capabilities must feature prominently as well in the coming years, as VLBI pulsar astrometry will be in demand for population studies that harvest the results from a number of SKA pathfinders [5][2]. Finally, it has been demonstrated that VLBI offers very interesting scientific applications for planetary Space missions, by providing accurate positions of spacecraft. Here too, real time processing is an important asset for the VLBI system of the future[6].

In all the science areas the VLBI sensitivity must be improved to maintain the EVN's complementary role with the EVLA and e-MERLIN facilities, by providing high-resolution diagnostics at a similar sensitivity. Besides increases in bandwidth, sensitivity improvements are also expected from expanding the number of antennas that participate in VLBI, as various new antennas are planned in Europe and beyond. A next generation correlator must provide capacity for processing these higher bandwidths, but also more flexibility for spectral line science, astrometry, pulsar binning and many bit processing.

In the long term, it should be noted that very long baseline capabilities are not completely covered by the SKA design, certainly not in its first two phases. Therefore, there will have to be a focus on the use of Global baselines and frequencies of 5 GHz and higher. Building on the existing international collaborations in Global VLBI and NEXPReS[3] (Novel Explorations Pushing Robust e-VLBI Services), this array could continue to have coverage in the Northern hemisphere and encompass the current EVN telescopes. A more flexible global VLBI array could be envisioned that can be deployed in tailor-made configurations for the best science.

## 3. Progress on e-VLBI

### 3.1 EXPReS programme

In the last decade e-VLBI has evolved from an experimental technique, connecting a small number of telescopes at modest bandwidth into an operational astronomical service with competitive sensitivity and imaging capabilities. The initial project has been supported by the EC through the FP6 Integrating activity EXPReS, which has stimulated a large-scale collaboration between EVN technical staff and European Research Network providers.

The initial argument for developing e-VLBI has been the desire to use VLBI on transient phenomena, being able to access the data on variable sources on the timescale that they vary. Indeed, opening the parameter space accessible to VLBI has turned up some very interesting results and various new science themes have been presented [8].

However, the project has done more than enabling new science. It has also clearly demonstrated that e-VLBI is not a technique limited to do rapid response science. Because of its real-time nature it is also much more robust against failures which can be noticed immediately and addressed instantaneously. With dedicated connections this has proven to be true, even though the high-speed links potentially add an extra layer of complexity. Moreover, it has been demonstrated convincingly that e-VLBI can work for intercontinental baselines (Figure 2).

Besides reducing the delivery time to the astronomer, e-VLBI can also help in enhancing the flexibility of the observations by avoiding the complex logistics that are involved in shipping scarce recording media. Although the costs are currently considerable, in the long run it is expected that real-time connectivity will be cheaper and more environmentally friendly.

---

[3] NEXPReS is an Integrated Infrastructure Initiative (I3), funded under the European Union Seventh Framework Programme (FP7/2007-2013) under grant agreement n° RI-261525.

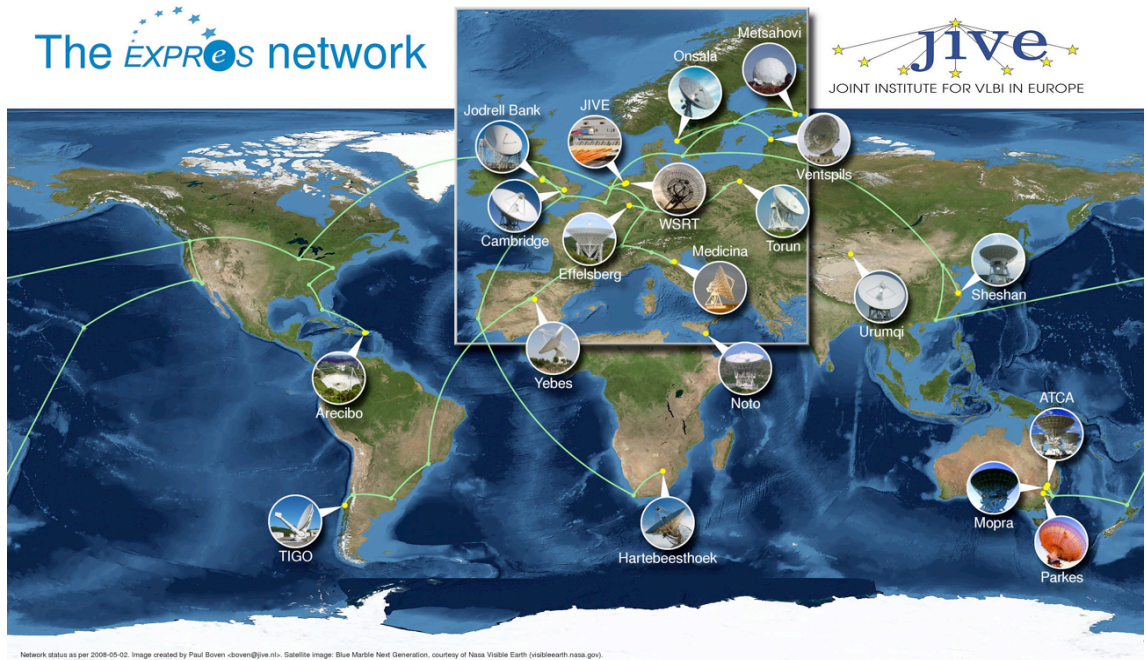

*Figure 2. Telescope and correlator configuration in the EXPReS network. Real-time connectivity to these telescopes was established during the course of the project.*

One way in which e-VLBI is less flexible than recorded VLBI, is that its current operations do not allow re-correlations, which are for example needed to accommodate large numbers of telescopes or spectral line experiments that need mixed bandwidths. It is noted that in principle all of these are limitations of the current correlator, but even so it will be advantageous to introduce buffering of the VLBI data in various stages in order to combine the best aspects of recorded and real-time operations.

**3.2 NEXPReS programme**

The next objective in the development of the VLBI of the future is to implement an e-VLBI component for every VLBI experiment that is carried out. By implementing transparent buffering mechanisms at telescope and correlator end, one can address all the current and future bottlenecks in e-VLBI. We envision an operational model in which all data that can be transported in real-time are correlated on the fly. Ideally such an e-VLBI experiment is completed directly after the observation, but there can be various logistic, technical and even scientific reasons, to access the data that was buffered and process the experiment a second time, possibly based on the evaluation of the real-time output. This way one could overcome limited connectivity to an essential station, or failures in streaming or correlation. Also reprocessing with different correlation parameters or new astronomical information on, for example, source position would be possible. Such a scheme would also end almost all transport needs of the physical recordings, as buffering capacity on both ends could accommodate all relay scenarios. Implementation of this service requires high-speed parallel recording hardware (many Gbps), as well as software systems that hide the bookkeeping from the VLBI components and their operators.

Besides these transparent, high bandwidth caching mechanisms, NEXPReS also has work-programmes on Bandwidth-on-Demand protocols. With these in place, it should be possible to reserve guaranteed, high-capacity connections for VLBI experiments and release them back for other applications afterwards. Also the distributed correlator effort focuses on allocating resources flexibly. Using computing resources at the astronomy sites it should be possible to farm out the correlation needs for a modest VLBI network deployed for transient event follow-up.

## 4. Technology development

While the full 1024Mbps data-rate has been offered regularly for e-VLBI observations. The EVN has started to push the sensitivity by testing with 4 Gbps in 2012. A critical item has been the introduction of the new filters and digitizers, the DBBC system in the case of the EVN. This system is needed for any new telescope that joins the array and eventually for all stations to go beyond 1024Mbps. Moreover, the EVN bandwidth is limited by the inherent width of the IF systems of the telescopes, and this must be addressed at many stations in the future. Eventually, there is the ambition to go to even higher bandwidths, which will be mostly useful at frequencies above 10 GHz. Here possibly up to 2 GHz bandwidth could be realized in each polarization, calling for at least 16 Gbps data-rates.

The future use also dictates that the existing VLBI arrays become much more flexible in scheduling and operations. Operational models must evolve in such a way that users can easily define the optimal array in extent, frequency and sensitivity for their observations, even on short notice. It should be possible to flexibly schedule and adapt observations in order to reach a guaranteed level of quality, taking for example atmospheric conditions into account.

### 4.1 Next Generation correlator

The EVN Mk4 correlator at JIVE that has been at the heart of the EVN operations since its dedication in 1998 is capable of processing 16 stations at 1024 Mbps. Although yet only a modest number of experiments exceed these parameters, the processing is now predominantly carried out on the EVN software correlator at JIVE (SFXC), which implements a more accurate correlation algorithm in a more flexible architecture. The SFXC platform offers a number of additional new scientific capabilities, including pulsar binning, wide field imaging and high spectral resolution modes. In particular it is very powerful for near-field correlation to measure spacecraft signals. This technique could be useful to help the correlation of Space VLBI data such as from the current RadioAstron mission [7][3][10].

For the longer-term future a large, new EVN data processor is required. Looking at the above specifications and the EVN (and global) ambitions to have more stations, one can anticipate the need to process 32 stations (a factor 4 in baselines compared to the current), 4 or maybe even 16 Gbps per second (a factor 4) and maybe a 4 to 8 times better spectral resolution than the current EVN data processor. Overall the aim must be to develop a correlator that is close to a hundred times more powerful than the current data processor. These are similar specifications as the correlators for SKA precursors (MeerKAT, ASKAP), and some of the other pathfinders (e-MERLIN, EVLA, APERTIF). Taking future energy constraints into consideration, JIVE is actively pursuing for correlation on FPGA based architectures (Figure 3); a similar conclusion as was reached by other SKA pathfinder projects.

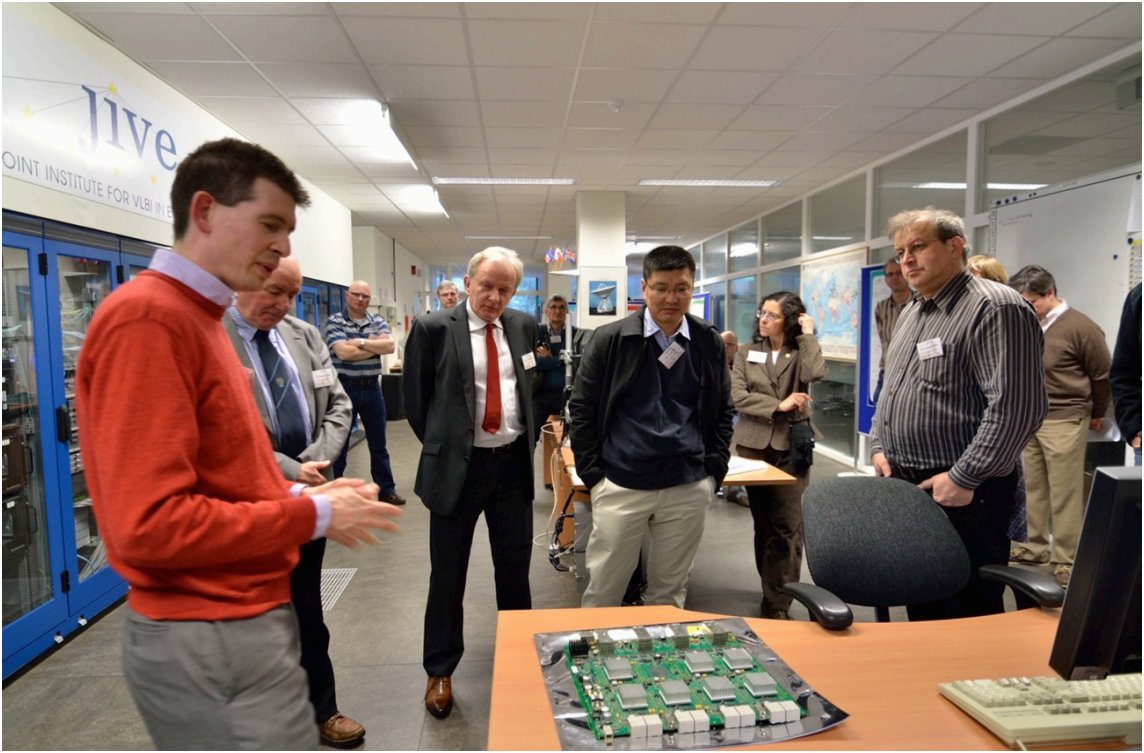

*Figure 3. Demonstration of the UniBoard prototype during the review of JIVE on March 5 2012*

### 4.2 Telescopes

One of the main motivations to increase the correlator capacity is the increasing number of telescopes. In global observations the number of telescopes can already go beyond 16. In the near future it will be possible to observe with EVN+MERLIN, using all stations. Recently first fringes were obtained with the Latvian Ventspils 32m telescope and the Sardinia Radio Telescope (64m) is reaching completion. In China the new telescopes of Miyun (50m) and Kunming (40m) have participated in experiments while a large new antenna is being erected in Shanghai [4]. In addition to newly constructed telescopes, the Russian telescopes in the KVASAR network have recently joined the EVN and first fringes have been obtained between Korean and European antennas.

A very interesting option to enhance the VLBI capacity is the project to retrofit existing communication dishes for astronomical research, as they become obsolete with the advent of fibre cables around the world. The possibilities include telescopes in geographically interesting locations such as the Goonhilly station in Cornwall and the Azores and Madeira. In particular, a large number of antennas have been identified around the African continent [5]. It has been recognized that establishing an African VLBI Network offers not only very good scientific capabilities, but also is a great opportunity for building technological and scientific capacity in Africa.

## 5. Conclusions

All these new dishes will contribute to the sensitivity and the imaging quality, while the global distribution allows for more and longer coverage during observations. Moreover, some of the new antennas are extending the high frequency capability of the EVN in the 22 and 43 GHz range. With such a facility in place, the VLBI of the future could also be much more flexible and ready to observe at any time with the most optimal array, never compromising the optimal scientific return, yet providing immediate feedback as well. This is the VLBI facility that future users want to see in order to follow-up SKA (pathfinder) results.

However, it is a challenge to shape such a Global VLBI ambition in parallel with the definition and construction of the SKA. It is important to continuously demonstrate that the VLBI facilities can continue to play an important role for training and education, as well as enable cutting edge astronomy research in Europe during the SKA engineering phases. Moreover, the national VLBI telescopes have a unique role in keeping the public aware of the fascinating science that can be done with radio astronomy. All the above aspects have been recognised, along with the technical developments below, in a recent review of JIVE. It was concluded that the institute is excellently positioned for the future, especially when it becomes a formal European Research Infrastructure Consortium (ERIC). With all these policy, technical and scientific ingredients in place, we can continue thankfully to *Resolve The Sky*.

## References

[1] R. Booth, *The beginnings of the EVN, JIVE and early space VLBI in Europe,* PoS(RTS2012)005

[2] P.J Diamond, *Australian SKA Pathfinder Telescope,* PoS(RTS2012)032

[3] L. Gurvits, *Space radio astronomy in the next 1000001 (binary) years,* PoS(RTS2012)045

[4] X.Y. Hong, *The developement of VLBI in China and its related to EVN,* PoS(RTS2012)012

[5] J. Jonas, *The MeerKAT Radio Telescope,* PoS(RTS2012)031

[6] D. Jones, *VLBI Astrometry of Planetary Orbiters*, PoS(RTS2012)046

[7] Y. Kovalev, *Space VLBI mission RadioAstron: current status and early science program,* PoS(RTS2012)040

[8] Z. Paragi, *SS433, microquasars, and other transients*, PoS(RTS2012)028

[9] K.L.J. Rygl, A. Brunthaler, A. Sanna, et al., 2012, A&A 539, 79

[10] R. Schilizzi, *The Story of Space VLBI,* PoS(RTS2012)010

[11] R. Vermeulen, *Resolving The Sky with LOFAR - Status and Results,* PoS(RTS2012)035

[12] J. Womersley, *SKA - current status,* PoS(RTS2012)033